\documentclass[nojss,noheadings]{jss}

\usepackage{dsfont}
\newcommand{\I}{\mathds{1}}

\usepackage{amssymb, amsmath,amsfonts}

\newcommand{\bfx}{\mathbf{x}}

\newcommand{\theoy}{Y}
\newcommand{\theox}{\mathbf{X}}
\newcommand{\theoz}{\mathbf{Z}}
\newcommand{\theodat}{(\theoy, \theox)}

\newcommand{\spacz}{\mathcal{Z}}
\newcommand{\obsy}{y}
\newcommand{\obsx}{\mathbf{x}}
\newcommand{\obsz}{\mathbf{z}}
\newcommand{\obsdat}{(\obsy, \obsx)}

\newcommand{\model}{\mathcal{M}}

\newcommand{\bfalpha}{\boldsymbol{\alpha}}
\newcommand{\bfbeta}{\boldsymbol{\beta}}
\newcommand{\bfgamma}{\boldsymbol{\gamma}}
\newcommand{\bftheta}{\boldsymbol{\vartheta}}

\newcommand{\bfsigma}{\boldsymbol{\sigma}}
\newcommand{\allparm}{\bfalpha, \bfbeta, \bfgamma, \bfsigma}
\newcommand{\htheta}{\hat{\bftheta}}
\newcommand{\halpha}{\hat{\alpha}}
\newcommand{\hbeta}{\hat{\beta}}

\newcommand{\alphaz}{{\bfalpha (\obsz)}}
\newcommand{\betaz}{{\bfbeta (\obsz)}}

\newcommand{\segb}{\mathcal{B}}

\newcommand{\sumb}{\sum\limits_{b=1}^B}
\newcommand{\sumi}{\sum\limits_{i=1}^N}

\newcommand{\alsfrs}{\text{ALSFRS}}

\DeclareMathOperator*{\argmin}{arg\,min}
\DeclareMathOperator*{\mE}{\mathbb{E}} 
\DeclareMathOperator*{\mP}{\mathbb{P}}
\DeclareMathOperator*{\normal}{\mathcal{N}}
\DeclareMathOperator*{\bernoulli}{\mathcal{B}}

\newcommand{\ie}{i.e.~}
\newcommand{\eg}{e.g.~}

\usepackage{multirow}
\usepackage[table]{xcolor}
\usepackage{tabularx, booktabs}

\usepackage{array}

\usepackage{listings}

\usepackage{tikz}
\usetikzlibrary{arrows,shapes}

\usepackage{caption}
\usepackage{subcaption}

\usepackage{rotating}

\usepackage{fancyhdr}
\fancypagestyle{plain}{
\fancyhf{}
\fancyfoot[LO,LE]{\small This is a preprint of an article accepted for publication in \textit{The International Journal of Biostatistics}.
  }
}

\title{Model-based Recursive Partitioning for Subgroup Analyses}

\author{Heidi ~Seibold\\ University of Zurich \And
Achim ~Zeileis\\ University of Innsbruck \And
Torsten ~Hothorn\\ University of Zurich}

\Plainauthor{Heidi Seibold, Achim Zeileis, Torsten Hothorn}

\Abstract{

The identification of patient subgroups with differential treatment effects
is the first step towards individualised treatments.  A current draft
guideline by the EMA discusses potentials and problems in subgroup
analyses and formulated challenges to the development of appropriate statistical
procedures for the data-driven identification of patient subgroups.  We
introduce model-based recursive partitioning as a procedure for the
automated detection of patient subgroups that are identifiable by
predictive factors.  The method starts with a model for the overall
treatment effect as defined for the primary analysis in the study protocol
and uses measures for detecting parameter instabilities in this treatment
effect.  The procedure produces a segmented model with differential treatment
parameters corresponding to each patient subgroup.  The subgroups are linked
to predictive factors by means of a decision tree.  The method
is applied to the search for subgroups of patients suffering from amyotrophic
lateral sclerosis that differ with respect to their Riluzole treatment
effect, the only currently approved drug for this disease.

}


\Address{

Heidi~Seibold\\
Department of Biostatistics\\
Epidemiology, Biostatistics and Prevention Institute\\
University of Zurich\\
Hirschengraben 84\\
CH-8001 Zurich, Switzerland\\

Achim Zeileis \\
Department of Statistics \\
Faculty of Economics and Statistics \\
University of Innsbruck \\
Universit\"atsstr. 15 \\
A-6020 Innsbruck, Austria \\

Torsten Hothorn \\
Department of Biostatistics\\
Epidemiology, Biostatistics and Prevention Institute\\ 
University of Zurich\\
Hirschengraben 84\\
CH-8001 Zurich, Switzerland\\
Email: \texttt{Torsten.Hothorn@uzh.ch}

}

\begin{document}

\maketitle



\section{Introduction}

With the rise of personalised medicine, the search for individual treatments
poses challenges to the development of appropriate statistical methods. 
Subgroup analyses following a traditional statistical assessment of an
overall treatment effect of a new therapy aim at 
identifying three groups of patients: (1) those who benefit from the new
therapy, (2) those who do not benefit, and (3) those whose clinical
outcome under the new therapy is worse than under alternative therapies. 
Such post-hoc subgroup analyses potentially lead to better benefit-risk
decisions and treatment recommendations but are subject to all kind of
biases and can hardly be performed under full statistical error control.
Therefore, the European Medicines Agency (EMA) recently published a draft of a guideline for
the investigation of subgroups in confirmatory clinical trials
\citep{ema_ema_2014} that discusses potential areas of application, 
necessity, pitfalls, and good practice in subgroup analyses.
In the guideline draft, three scenarios in which exploratory investigation
of subgroups is of special interest were identified:
\begin{description}
  \item[Scenario 1:] ``The clinical data presented are overall
statistically persuasive with therapeutic efficacy demonstrated globally. 
It is of interest to verify that the conclusions of therapeutic efficacy
(and safety) apply consistently across subgroups of the clinical trial
population.''
	\item[Scenario 2:] ``The clinical data presented are overall
statistically persuasive but with therapeutic efficacy or benefit/risk which
is borderline or unconvincing and it is of interest to identify post-hoc a
subgroup, where efficacy and risk-benefit is convincing.''
	\item[Scenario 3:] ``The clinical data presented fail to establish
statistically persuasive evidence but there is interest in identifying a
subgroup, where a relevant treatment effect and compelling evidence of a
favourable risk-benefit profile can be assessed.''
\end{description}
Especially in trials with highly heterogeneous study populations, 
subgroup analyses can help to reduce the variability of the estimated 
overall treatment effect by splitting the study population into more
homogeneous subgroups.

Information about the individual treatment effect might be available from
cross-over trials or from counterfactual analyses of parallel-group designs
\citep{holland_statistics_1986, gadbury_unittreatment_2000}.  These individual
effects can then be linked to potentially predictive variables. In the absence
of such information, most importantly in the case of parallel-group designs
studied here, subgroup analyses can be seen as the search for or specification
of treatment $\times$ covariate interactions and we proceed along this path.
A covariate measures a patient characteristic
that potentially explains the patient's individual treatment effect.  In the
commonly applied models with linear predictors, such as the linear,
generalised linear or linear transformation models, the specification of
higher-order interaction terms and especially the subsequent inference are
known to be burdensome.  For non-categorical covariates, it is a priori
unclear how one can derive a subgroup from a significant treatment $\times$
covariate interaction.

Automated interaction detection \citep{morgan_problems_1963}, today known as
recursive partitioning methods or simply ``trees'', was suggested as an
interaction search procedure more than $50$ years ago, and has had a very active
development community ever since. Although the application of trees for
subgroup identification seems to be straightforward, no generally applicable
method is available \citep{doove_comparison_2014}. The main technical problem is
that classical trees were developed for identifying higher-order
covariate interactions but additional work is required to restrict interactions
to treatment $\times$ covariate interactions. Due to the non-parametric
nature of most tree models, blending trees with the linear models typically used
to describe the treatment effect is challenging.

While setting up such automated procedures for subgroup identification, one has
to bear in mind that the impact of a covariate on the endpoint can be
prognostic, predictive, or both. Prognostic factors have a direct impact on the
endpoint, independent of the treatment applied. This corresponds to a main
effect.  A predictive factor explains a differential treatment effect, \ie a
treatment $\times$ covariate interaction term .  Both the main and the
treatment interaction terms are important for factors that are prognostic and
predictive at the same time \citep{italiano_prognostic_2011}.

In our analysis, we aimed at detecting subgroups of patients suffering from
amyotrophic lateral sclerosis (ALS) in which the subgroups differ in the
effect of treatment with Riluzole, the only approved drug for ALS treatment
today.  The two endpoints of interest are a functional endpoint assessing
the patient's ability to handle daily life and the overall survival time. 
We estimated the overall treatment effect of Riluzole using four different
base models; the choice of the model depended on the measurement scale of
the endpoint.  A normal generalised linear model (GLM) with log-link was
used for the sum-score of the functional endpoint, and item-specific
proportional odds models were used for the decomposed score.  For the
right-censored survival times, we used a parametric Weibull model and a
semiparametric Cox model.  Our aim was to partition these linear models with
respect to the treatment effect parameter and to develop a segmented model
that includes treatment $\times$ covariate interactions that describe the
relevant subgroups.

We applied model-based recursive partitioning
\citep{zeileis_model-based_2008} to the functional and survival models
describing the effect of Riluzole on ALS patients in order to obtain subgroups
with a differential treatment effect.  The main advantage of
embedding our subgroup analysis into this general framework of model
partitioning is that one can partition the base model used for analysing the
overall treatment effect, regardless of the measurement scale of the
endpoint.  The method allows us to focus attention on predictive factors, while
other terms, such as the effects of strata or nuisance parameters, can be
held fixed. 

Section~\ref{mob} introduces the general framework for subgroup
identification and compares the new procedure to methods published
previously in the light of this general theoretical framework.  In Section~\ref{applic}, 
we present results of our subgroup analysis of Riluzole treatment of ALS
patients and discuss the patient subgroups and corresponding differential
treatment effects found.

\section{Model-based recursive partitioning for subgroup identification}\label{mob}

Subgroup analyses require the definition of a parameter describing the
treatment effect.  In clinical trials, this parameter is typically already
contained in the model that was defined in the study protocol for the analysis
of the primary endpoint.  The treatment effect was estimated in the primary
analysis under the assumption that the corresponding parameter is universally
applicable to all patients.  In the presence of subgroups, this assumption does
not hold and these patient subgroups differ in their treatment
effect.  If we assume that the different treatment effects can be understood as
a function of patient characteristics, the patient subgroups
can be identified by estimating this treatment effect function.  Model-based
recursive partitioning can be employed as a procedure for the estimation of
such a treatment effect function and the identification of the corresponding
patient subgroups. The name of the procedure comes from the nature of the
algorithm that recursively partitions the initial model used for the analysis of the
primary endpoint.\\

\subsection{Model and algorithm}

We started with a model $\model(\theodat, \bftheta)$ that describes the
conditional distribution of the primary endpoint $Y$ (or certain
characteristics of this distribution) as a function of the treatment arm
and potentially further covariates (both contained in $\theox$) through
parameters $\bftheta$ as defined in the study protocol.  The parameter
vector $\bftheta = (\allparm)^\top$ typically contains one or more intercept
parameters $\bfalpha$, one or more treatment effect parameters $\bfbeta$,
other model parameters $\bfgamma$, \eg\ effects of covariates, and
potential nuisance parameters $\bfsigma$, \eg\ the error variance in a linear
model.  The estimator is defined as the minimizer of an objective function
$\Psi$, which usually is the negative log-likelihood:
\begin{align}\label{esttheta}
  \htheta &= \argmin\limits_\vartheta \sumi \Psi(\obsdat_i, \bftheta).
\end{align}
Estimating $\bftheta$ is equivalent to solving the score equation
\begin{align}
	\sumi \frac{\partial \Psi(\obsdat_i, \bftheta)}{\partial \bftheta} =
	\sumi \psi(\obsdat_i, \bftheta) &= 0,
\end{align}
where $\psi$ is the score function, \ie the gradient of the objective
function $\Psi$ with respect to $\bftheta$. The model framework is more
general than the log-likelihood framework because $\Psi$ is not necessarily
a negative log-likelihood function.

In the presence of patient subgroups that differ in their
treatment effect $\bfbeta$, an estimate $\hat{\bfbeta}$ obtained for all
patients $i = 1, \dots, N$ in the study only reflects the mean treatment
effect but ignores that the success or failure of a specific treatment might
depend on additional characteristics of each individual patient.  We describe patient
subgroups as a partition $\{\segb_b\}$ ($b=1,\dotsc,B$) of all patients $i = 1,
\dots, N$. The subgroup-specific model parameters are then $\bftheta(b)$.
These parameters can in general be seen as varying coefficients
\citep{hastie_varying-coefficient_1993}, however they may depend on several
patient characteristics and are always step functions with a different level for
each subgroup and not only a smoothly varying coefficient for one single
predictive variable.

Since we are searching for predictive and prognostic factors, we are only
interested in subgroups that differ in the intercept or the treatment effect or
both as explained in Section~\ref{cont_int}. With $\bftheta(b) = (\bfalpha(b), \bfbeta(b), \bfgamma,
\bfsigma)^\top$ we assume that the effects of covariates and nuisance
parameters are constant for all patients.
The partition $\{\segb_b\}$ is defined by $J$ partitioning    variables $\theoz
= (Z_1,\dotsc,Z_J) \in \spacz$; in other words,   $\{\segb_b\}$ is a hypercube
in the $J$-dimensional sample space $\spacz$.  These partitioning variables
$\theoz$ are the additional patient characteristics that potentially influence
$\bfalpha(b)$ and $\bfbeta(b)$.  If for example gender were a predictive factor
in a given treatment-endpoint relationship, it would be a patient
characteristic that is involved in forming the partitions.
If the partition $\{\segb_b\}$ is known, the partitioned model parameters
$\bftheta(b)$  could be estimated by minimising the segmented objective
function:
\begin{align}\label{esttheta1}
	(\htheta(b))_{b = 1,\dots,B} &= \argmin\limits_{\bftheta(b)} \sumi
	\sumb \I\left( \obsz_i \in \segb_b \right) \Psi(\obsdat_i,
	\bftheta(b)),
\end{align}
where $\I$ denotes the indicator function and $\obsdat_i, \obsz_i$ are the 
realisations of $\theodat$ and $\theoz$ for the $i$-th patient.
This allows us to write the subgroup-specific intercept and treatment
parameters as a function of the partitioning variables
\begin{align*}
	\alphaz = \sumb \I(\obsz \in \segb_b) \cdot \bfalpha(b)
	\quad\text{ and }\quad
	\betaz = \sumb \I(\obsz \in \segb_b) \cdot \bfbeta(b).
\end{align*}

Without any a priori knowledge about the partition $\{\segb_b\}$, we 
want to estimate the functions $\alphaz$ and $\betaz$ by means of
model-based recursive partitioning.  The main idea underlying this method is
the ability to detect parameter instabilities, \ie non-constant parameters
in a parametric or semiparametric model, by looking at the score function.
Because we are only interested in detecting non-constant intercepts
$\alphaz$ and treatment effects $\betaz$, we focus on the partial score
functions $\psi_\alpha(\theodat, \bftheta) =
 \partial \Psi(\theodat, \bftheta)/\partial \bfalpha$
and 
$\psi_\beta(\theodat, \bftheta) = \partial
 \Psi(\theodat, \bftheta)/\partial \bfbeta$.
If the model parameters are in fact constant and do not depend on any
of the partitioning variables $\theoz$, the partial score functions
$\psi_\alpha(\theodat, \bftheta)$ and $\psi_\beta(\theodat, \bftheta)$
are independent of $\theoz$. Consequently, parameter instability
corresponds to a correlation between either of the partial score functions
and at least one of the partitioning variables $Z_1, \dots, Z_J$.
In order to formally detect deviations from independence between the
partial score functions and the partitioning variables, model-based 
recursive partitioning utilises independence tests. The null hypotheses
\begin{align*}
	H_0^{\alpha,j}&: \quad \psi_\alpha(\theodat, \htheta) \quad\bot\quad Z_j, j = 1, \dots, J\\
	&\text{and}  \\
	H_0^{\beta,j}&: \quad \psi_\beta(\theodat, \htheta) \quad\bot\quad Z_j, j = 1, \dots, J
\end{align*}
for a given model $\model(\theodat, \htheta)$ state that the partial score
functions with respect to $\bfalpha$ and $\bfbeta$, respectively, are independent
of the partitioning variable $Z_j$ ($j = 1, \dots, J$). Hence, these null 
hypotheses correspond to an appropriate model fit regarding the intercept and treatment 
parameter. 
Because the partial score functions under the null hypotheses are at least
asymptotically normal in many model families, asymptotic M-fluctuation tests
with appropriate correction for multiplicity were introduced for model-based
recursive partitioning by Zeileis and coworkers
\citep{zeileis_generalized_2007, zeileis_model-based_2008}.  
Alternatively, permutation tests can be
applied in situations where asymptotic normality of the partial score is not
guaranteed \citep{zeileis_toolbox_2013} or in cases with small numbers of
observations \citep{hothorn_implementing_2008, hothorn_lego_2006,
hothorn_unbiased_2006}, which are common in medicine. 
Also in this case procedures for multiple testing are used to cope with
a possibly large number of partitioning variables $J$.

If we can reject at least one of the $2 \times J$ null hypotheses for the global model 
$\model(\theodat, \htheta)$ at a pre-specified nominal level, model-based
recursive partitioning selects the partitioning variable $Z_{j^\star}$
associated with the highest correlation to any of the partial score
functions.  This is usually done by means of the smallest $p$-value.  The
dependency structure between the partitioning variable $Z_{j^\star}$ and
either one of the partial score functions is described by a simple cut-point
model.  Once we find an optimal cut-point $Z_{j^\star} < \mu$ using a
suitable criterion \citep{zeileis_model-based_2008, hothorn_unbiased_2006},
we split the patients into two subgroups according to $Z_{j^\star} < \mu$. 
For both subgroups, we estimate two separate models with parameters 
$\htheta(1)$ and $\htheta(2)$, respectively, obtain the corresponding
partial score functions, and test the independence hypotheses.  If we find
deviations from independence, we in turn estimate a cut-point in the most
highly associated partitioning variable, and split again.  The procedure of
testing independence of partial score functions and partitioning variables
is repeated recursively until deviations from independence can no longer be
detected. 

Since model-based recursive partitioning is a tree method, in the following
we use topic-specific vocabulary, such as nodes. The root node contains
all patients and is the basis for the initial model, inner nodes represent splits
and leaf nodes contain the patients of the different subgroups and specify
the partition-specific models. The paths from root to leaf nodes define the subgroups.

\subsection{Content interpretation}\label{cont_int}

A clearer picture of the interpretation of subgroup-dependent model
parameters and distribution of the partial scores under unstable parameters
is best given by means of a partitioned linear model discussed in the
following.

Here $x_{A}$ is a contrast that indicates whether a subject was treated with
treatment $A$ (active) but not $C$ (control) in a two-armed trial and
$x_{\text{stratum}}$ is a stratum with $\obsx = (x_A, x_{\text{stratum}})$.
The conditional distribution of the primary endpoints $Y$ given treatment and
stratum is normal
\begin{align}
  \theoy|\theox = \obsx &~\sim ~ \normal(\alpha + \beta x_{A} + \gamma
x_{\text{stratum}},~ \sigma^2).
\end{align}
The segmented model we want to fit using model-based recursive partitioning reads
\begin{align}
	\theoy|\theox = \obsx, \theoz = \obsz &~\sim ~
\normal(\alpha(\obsz) + \beta(\obsz) x_{A} + \gamma
x_{\text{stratum}},~ \sigma^2),
\end{align}
where $\gamma$ is the effect of the stratum and the variance $\sigma^2$ is a 
nuisance parameter. The objective function for a patient with observations
$\obsdat$ is the negative log-likelihood, when maximum likelihood estimation is
used, or the error sum of squares, when ordinary least squares is used. Yet,
both methods lead to the same scores
\begin{align}
	\psi(\obsdat, \htheta) &= \left(
		\begin{array}{l}
		\left.\frac{\partial \Psi(\obsdat, \theta)}{\partial \alpha}\right|_{\theta = \htheta}\\
		\left.\frac{\partial \Psi(\obsdat, \theta)}{\partial \beta}\right|_{\theta = \htheta}\\
		\end{array} 
	\right)^{\top} = \frac{1}{\sigma^2}\left(
		\begin{array}{l}
		y - (\halpha + \hbeta x_{A} + \hat{\gamma} x_{\text{stratum}})\\
		(y - (\halpha + \hbeta x_{A} + \hat{\gamma}
		x_{\text{stratum}})) \cdot x_{A}\\
		\end{array} 
	\right)^\top
\end{align}
and thus to the same solution. Note that the partial score function with
respect to the intercept is proportional to the least-square residuals and
all further scores are proportional to the product of the residuals and the
respective variable.

\begin{sidewaysfigure}[p!]
\centering
\includegraphics[width=\textwidth]{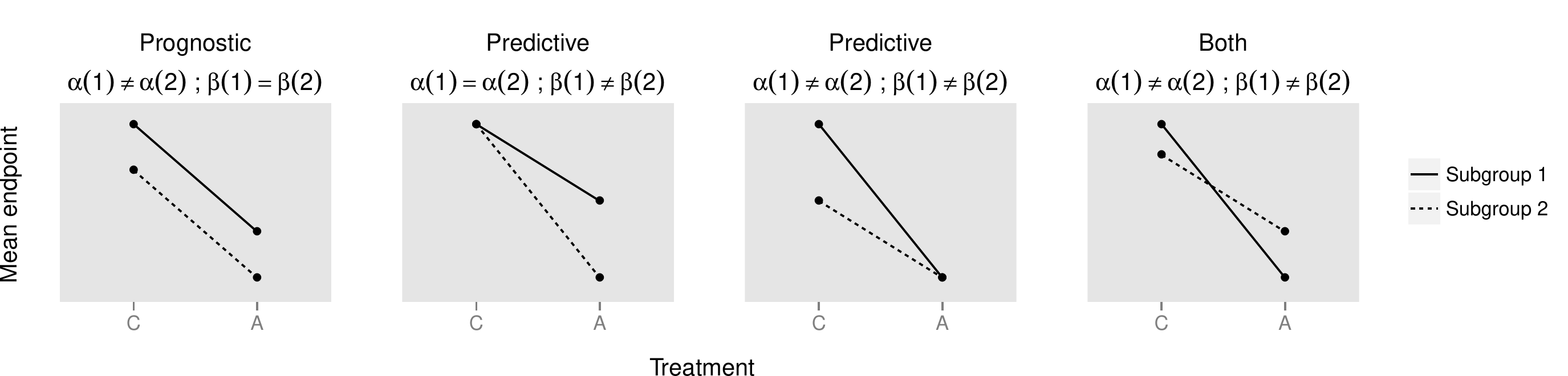}
\caption{Possible mean primary endpoint within subgroups resulting from a 
         predictive, prognostic, or predictive and prognostic variable.}
\label{fig.predprog}
\end{sidewaysfigure}

A partitioning variable can be predictive, prognostic, or both, and we have
to consider the parameters in the model to understand the nature of a
partitioning variable.  Figure~\ref{fig.predprog} shows examples for mean
primary endpoints and the corresponding intercept $\alpha$ and treatment
effect $\beta$.  If $\alpha(\obsz)$ varies over $\obsz$, but $\beta(\obsz)$
is constant, then the components of $\obsz$ are prognostic because the mean
primary endpoint varies but not the treatment effect (see first column of
Figure~\ref{fig.predprog}).  If $\beta(\obsz)$ varies over $\obsz$ and
$\alpha(\obsz)$ is constant, then the variables in $\obsz$ are predictive
since it means that the mean primary endpoint in one treatment arm stays
the same but the treatment effect changes over $\obsz$ (second column).  If
both parameters vary, then $\obsz$ is predictive (third column) or
predictive and prognostic at the same time (last column).  In the latter
situation, the mean primary endpoint of the second subgroup changes over
$\obsz$ and the intercept also changes.

\begin{figure}[htp]
\begin{subfigure}[b]{\textwidth}
\includegraphics[scale=0.6]{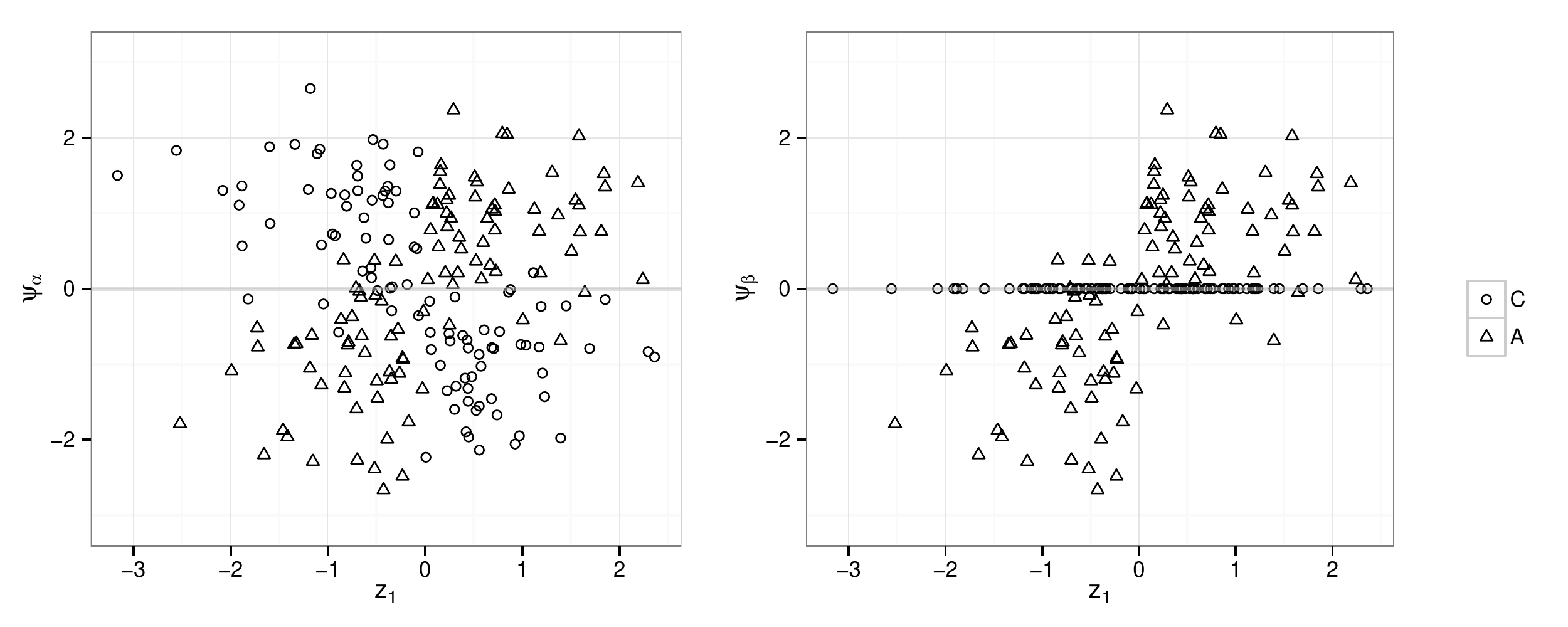}
\caption[Partial scores of a predictive and prognostic variable.]{Partial scores of a predictive and prognostic variable (equation \ref{mod.pred}).}
\label{fig.pred}
\end{subfigure}
\begin{subfigure}[b]{\textwidth}
\includegraphics[scale=0.6]{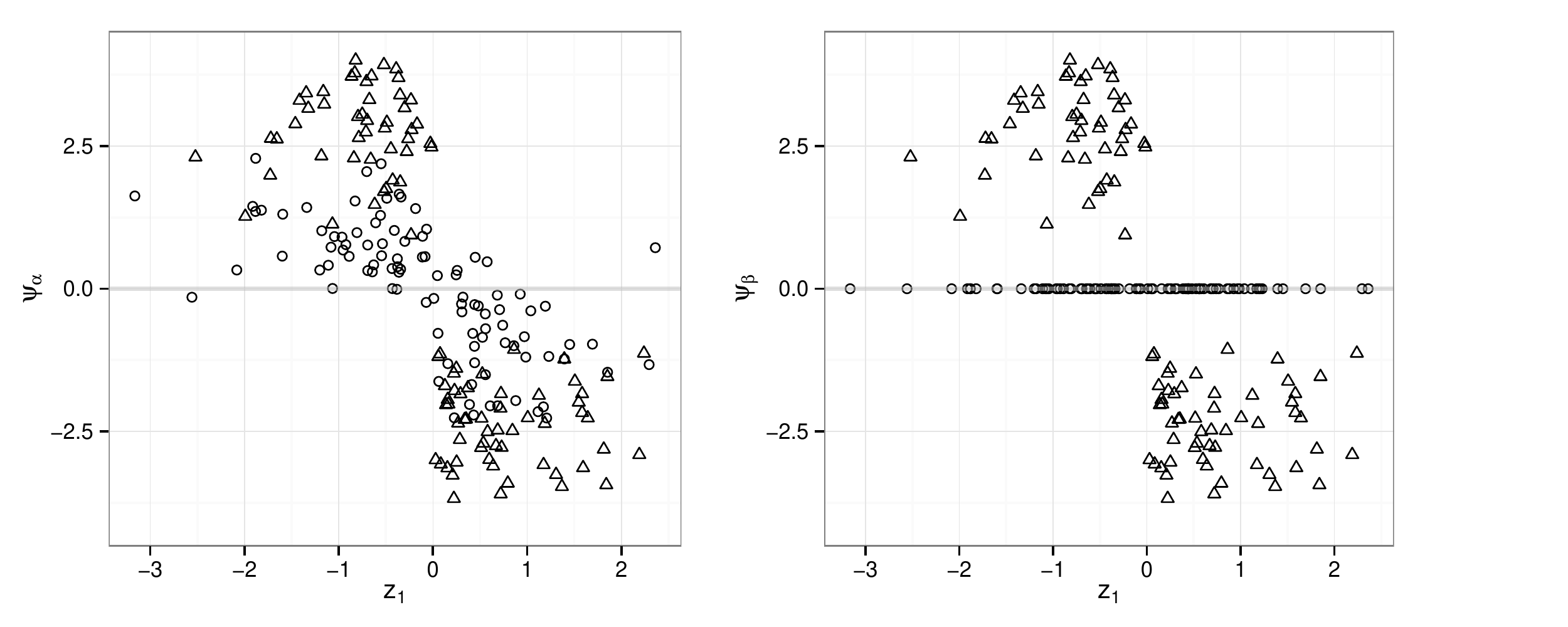}
\caption[Partial scores of a predictive and prognostic variable.]{Partial scores of a predictive and prognostic variable (equation \ref{mod.pred2}).}
\label{fig.pred2}
\end{subfigure}
\begin{subfigure}[b]{\textwidth}
\includegraphics[scale=0.6]{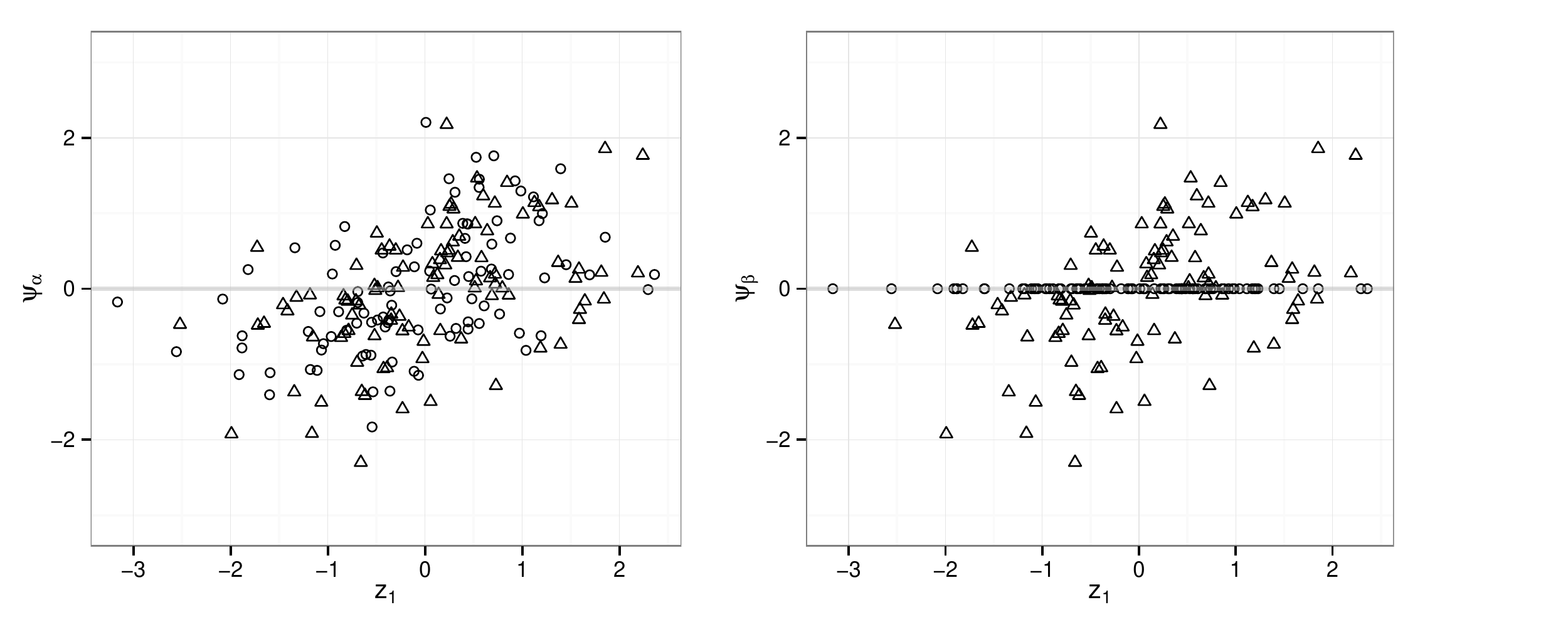}
\caption[Partial scores of a prognostic variable.]{Partial scores of a prognostic variable (equation \ref{mod.prog}).}
\label{fig.prog}
\end{subfigure}
\caption[Partial scores of different kinds of variables.]{Partial scores of different kinds of variables. The symbols represent the treatment arms C and A as indicated.}
\end{figure}

It is also interesting to take a closer look at the partial scores. 
Figure~\ref{fig.pred} shows the partial scores with respect to intercept and
treatment parameter that result from a linear model $\theoy|\theox = \obsx \sim
\normal(\alpha + \beta x_{A},~ \sigma^2)$ plotted against a partitioning
variable $z_1$, which is predictive and prognostic.  The data-generating
process of this model was suggested by \cite{loh_regression_2014} and is defined as
\begin{align}
	\theoy|\theox = \obsx, \theoz = \obsz &~\sim~ \normal(1.9 + 0.2 \cdot x_{A}  + 1.8 \cdot \I(z_1 < 0) + 3.6 \cdot \I(z_1 > 0) \cdot x_{A} , ~0.7),
  \label{mod.pred}
\end{align}
with $X_A$ from $\bernoulli(1, 0.5)$ and $Z_1$ from $\normal(0, 1)$. For the
example, we used this process to draw a sample of 200 observations.

The partial scores with respect to the intercept $\psi_\alpha$ 
fluctuate randomly around zero over the whole range of $z_1$. 
The partial scores with respect to the treatment parameter $\psi_\beta$ 
change. Hence, in this situation, model-based recursive partitioning 
would detect a deviation from independence between $\psi_\beta$ and $z_1$ 
and implement a split at approximately $z_1 < 0$. There is no chance of
finding this cut-point by looking at the least-square residuals only, since
a deviation of independence between $\psi_\alpha$ and $z_1$ is hardly
visible in the scatterplot in the left panel of Figure~\ref{fig.pred}.
Figure~\ref{fig.pred2} shows the partial scores obtained with a slightly modified 
data-generating process, where instead of $\I(z_1 > 0) \cdot x_{A}$, one has $\I(z_1
< 0) \cdot x_{A}$:
\begin{align}
  \theoy|\theox = \obsx, \theoz = \obsz &~\sim~ \normal(1.9 + 0.2 \cdot x_{A}  + 1.8 \cdot \I(z_1 < 0) + 3.6 \cdot \I(z_1 < 0) \cdot x_{A} , ~0.7).
  \label{mod.pred2}
\end{align}
Here the procedure would split the partial score with 
respect to the intercept, although $z_1$ is still prognostic and predictive
at the same time.

If we focus on the prognostic variable $z_1$ in the model
\begin{align}
	\theoy|\theox = \obsx, \theoz = \obsz &~\sim~ \normal(2 \cdot x_{A} + \I(z_1 > 0), ~0.7),
  \label{mod.prog}
\end{align}
we see non-random patterns in both scores (see Figure~\ref{fig.prog}). Since the partial scores with
respect to the treatment parameter are set to zero for treatment arm $A$,
we would split on basis of the scores with respect to the intercept, just as
a consequence of a higher power.

These three examples show that splitting in the partial score with respect
to the intercept does not give any information about whether the partitioning
variable is predictive or prognostic.  It also does not make sense to choose to
split only in the score with respect to the treatment parameter 
because one might miss important cut-points.  In order to be able to say whether
a partitioning variable is predictive or prognostic, it is not enough to know
which partial scores are responsible for the split.  It is necessary to consider
the model parameters in the segmented model.  If the treatment
parameter $\beta$ varies in the subgroups, then the chosen partitioning variables
are predictive or both predictive and prognostic.  If $\beta$ is constant, 
the variables are only prognostic.

\subsection{Relation to established procedures}

Traditional approaches for subgroup identification are also based on a model
for the primary endpoint, but the segmentation is implemented by means of
varying coefficients. More precisely, the model includes interactions between
treatment and the patient characteristics $\obsz$ in addition to the main
effects 
\begin{align} 
  \mE(\theoy|\theox = \obsx, \theoz = \obsz) &= \alpha +
	\beta x_{A} + \bfgamma_\text{prognostic}^\top \obsz +
	\bfgamma_{\text{predictive}}^\top \obsz  x_{A}\\ &= (\alpha_0 +
	\bfgamma_{0,z}^\top \obsz) (1 - x_{A}) + (\alpha_1 +
	\bfgamma_{1,z}^\top \obsz) x_{A}, \nonumber 
\end{align} 
with $\alpha = \alpha_0,~ \beta = \alpha_1 - \alpha_0,~
\bfgamma_{\text{prognostic}}^\top =
\bfgamma_{0,z}^\top~$ and $\bfgamma_{\text{predictive}}^\top =
\bfgamma_{1,z}^\top -  \bfgamma_{0,z}^\top$.
The model is known as the ``classical approach'' for subgroup analyses 
\citep{kehl_responder_2006, foster_subgroup_2011}. 
Significant interaction terms $\bfgamma_\text{predictive}$ are in this case subject to 
the choice of relevant partitioning variables. However, patient subgroups can
only be identified directly in this model for categorical variables $z_j$
since the model has no notion of optimal cut-off points. As the number of
potential partitioning variables $J$ might be large, the simultaneous
estimation of all parameters in the model might be computationally
burdensome and associated with a large variance. Regularisation procedures
may be applied for selecting relevant interaction parameters that deviate
considerably from zero.

RECPAM \citep{ciampi_tree-structured_1995, negassa_tree-structured_2005} goes 
a step further and fits such models by trees. In every node, a likelihood-ratio 
test is computed that compares the segmented model
\begin{align}
	\mE(\theoy|\theox = \obsx, \theoz = \obsz) = 
             \alpha + \beta_1  x_{A} \I(z_j \in \segb_k) +  
                      \beta_2  x_{A} [1-\I(z_j \in \segb_k)]  
\end{align}
to the constant model 
\begin{align}
	\mE(\theoy|\theox = \obsx) = \alpha + \beta x_{A} 
\end{align}
for every possible segment $\segb_k$ ($k=1,\dotsc,K$) induced by all possible
cut-off points in $z_j$, \ie an exhaustive search is performed. 
The procedure is applied to all partitioning
variables $z_j$ ($j=1,\dotsc,J$). The algorithm then chooses the variable and 
segmentation that comes along with the highest test statistic. The method is so 
far limited to linear models and Cox proportional hazards models, and
parameter instabilities can only be detected in $\beta$ but not in
$\alpha$.

A method that is similar in spirit to model-based recursive partitioning
is the Gs method
\citep{loh_regression_2014} based on the GUIDE algorithm
\citep{loh_regression_2002, loh_improving_2009}.  Instead of using partial
scores with respect to intercept and treatment effect, Gs uses only the
least-square residuals (that is, only the partial score with respect to the
intercept).  In contrast to model-based recursive partitioning, Gs looks at
the dichotomised (at zero) residuals separately in the two treatment arms. 
The independency between positive/negative residual signs and each
partitioning variable is tested using a chi-squared test separately for
each treatment.  If the partitioning variable is at least ordinal, it is
dichotomised by splitting at the mean.  The optimal split variable 
chosen is the one that induces the highest sum of chi-squared statistics. 
Looking at the left panels of Figures~\ref{fig.pred} and \ref{fig.pred2}, one
can imagine that in these situations the procedure may successfully 
find the subgroups.  However, in a less clear situation and where the
optimal cut-point is not near the mean of $z_1$, the method will have
lower power or will not be able to find a split at all.

Another recently proposed tree algorithm is qualitative interaction trees \\
\citep[QUINT,][]{dusseldorp_qualitative_2013}.  QUINT searches for instabilities
in the treatment parameter $\beta$ only, but the resulting partitions have
to have different signs in the parameter.  In other words, QUINT aims at
finding subgroups in which the treatment effect is the reverse of that of the
other subgroups. 
The current implementation of QUINT \citep{dusseldorp_quint:_2013} is limited
to continuous primary endpoints.  It would be possible to enforce splits that
are qualitatively different in model-based recursive partitioning. This could
be achieved by incorporating a criterion that implements a split only if the
treatment effects in the two new subgroups have different signs.

SIDES \citep[subgroup identification based on differential effect search,][]{Lipkovich_Dmitrienko_Denne_2011}
and SIDEScreen \citep{Lipkovich_Dmitrienko_2014} aim at identifying subgroups of patients with 
high benefit from a novum compared to the standard treatment. Although the
subgroups are linked to hypercubes in the sample space of $Z$, they are 
overlapping and can therefore not be represented as a tree structure. The
methods are based on a cross-validated implementation of subgroups that were
obtained on independent learning samples.

More general approaches blending recursive partitioning with traditional models
\citep[known as hybrid, model, or functional trees in machine
learning,][]{gama_functional_2004} include M5 \citep{quinlan_C4.5_1993}, GUIDE
\citep{loh_regression_2002}, CRUISE \citep{kim_classification_2001}, LOTUS
\citep{chan_lotus_2004} and maximum likelihood trees \citep{su_maximum_2004}.
Bayesian approaches can be found in \cite{chipman_bayesian_2002} and
\cite{bernardo_bayesian_2003}.  Except GUIDE, none of these methods has been
studied in the specific context of subgroup analyses so far.

\section{Partitioning effects of Riluzole on ALS patients} \label{applic}

ALS is a neurodegenerative disease that
causes weakness, muscle waste and paralysis.  Currently the only drug on the
market for treating ALS is Riluzole (Rilutek). It slows down disease
progression but only modestly prolongs life expectancy by about two months
\citep{ema_Riluzole_2012}.  A more thorough
investigation of the treatment effect of Riluzole in ALS patients is of great
importance since a cure is not yet available and patients usually die within
$1.5$ to 4 years after disease onset \citep{chio_prognostic_2009}.
We use model-based recursive partitioning to address the question whether 
Riluzole has an especially low or high
treatment effect on both functional and survival endpoints of any subgroups of patients.

Our analysis is based on patient information obtained from the PRO-ACT
(Pooled Resource Open-Access ALS Clinical Trials) database
\citep{atassi_pro-act_2014}, which contains data of ALS
patients that were involved in one of several publicly- and
privately-conducted clinical trials.  The database provides information on
patient survival, functional endpoint (the ALS functional rating scale),
Riluzole use, demographics, family history, patient history, forced and slow
vital capacity, laboratory data and vital signs.  The data were fully
de-identified and therefore the centres of data ascertainment are not given
in the data set.  The participants gave their informed consent, and study
protocols were approved in the respective medical centres.  The database was
initiated by the non-profit organisation Prize4Life that aims at
accelerating cure and drug development for ALS, for example through
the DREAM-Phil Bowen ALS Prediction Prize4Life challenge \citep{Kueffner_Zach_Norel_2014}.

The ALS Functional Rating Scale \citep[ALSFRS,][]{brooks_amyotrophic_1996} is a
widely used instrument for evaluating the functional status of patients with
ALS even though the uni-dimensionality of the score seems questionable
\citep{franchignoni_evidence_2013}.  It is
a sum-score of the following ten items: speech, salivation, swallowing,
handwriting, cutting food and handling utensils, dressing and hygiene,
turning in bed and adjusting bed clothes, walking, climbing stairs, and
breathing.  Each of these items can have values from zero to four, where
four is normal and zero indicates the inability of performing the respective
action.  Hence, if the ALSFRS has a value $40$, the patient has 
normal abilities for all items.  The lower the score, the worse is the
patient's status.  The items were measured at several time points
during the study period.  We focused on the ALSFRS reading approximately six
months after treatment start as the functional endpoint.  Approximately means
that we used the measurement closest to six months after treatment start,
with a maximal absolute deviation of $20$ days.  In addition, we also
decomposed the score and modelled the items defining the score separately.

The survival time of patients was measured in days starting with the
patient's enrolment in one of trials.  For patients without survival
information, we used the latest follow-up time given for the patient in the
data as censoring time.  

Model-based recursive partitioning was applied to models for the functional
and survival endpoints.  We allowed parameter instabilities in both the
intercept and the Riluzole treatment effect.  Bonferroni-adjusted
permutation tests using test statistics of a quadratic form
\citep{hothorn_unbiased_2006} were applied for assessing independence of the
partial score functions and the partitioning variables and also for
cut-point selection.  The use of permutation tests for cut-point selection
improves speed compared to the original suggestion of fitting and comparing
models for all reasonable partitions \citep{zeileis_model-based_2008}.  We
restricted the depth of the trees to two levels. 
Parameter estimates including confidence intervals are given for the final
subgroups. Note that we are computing the confidence intervals after applying
model selection through splitting into subgroups and thus the intervals 
should be interpreted with caution.
For both endpoints, we
used partitioning variables available at patient enrolment from the
following groups of variables: demographics, family history, patient
history, forced and slow vital capacity, laboratory data, and vital signs. 
We excluded patient records with missing values at the endpoints; the sample
size was $N = 2534$ for the functional endpoint and $N = 3306$ with $916$
events for the survival endpoint.

\subsection{ALSFRS}

The ALSFRS six months after treatment start ($\alsfrs_6$) defined the functional endpoint.
The sum-score is positive, and the model needs to adjust for the
baseline ALSFRS obtained at treatment start ($\alsfrs_0$). We used
a normal GLM with log-link and offset $\log(\alsfrs_0)$ such that the model
\begin{align}
  \mE\left.\left(\frac{\alsfrs_6}{\alsfrs_0} \right\vert  X = x\right) =
\frac{\mE(\alsfrs_6\vert  X = x)}{\alsfrs_0} = \exp\{ \alpha + \beta x_{R}\}
\end{align}
describes the expected relative change in the ALSFRS over the first six
months under treatment. The treatment (Riluzole/no Riluzole) is indicated by
$x_R$. The model was fitted by maximum likelihood.

\begin{sidewaysfigure}
 \centering
 \includegraphics[width=\textwidth]{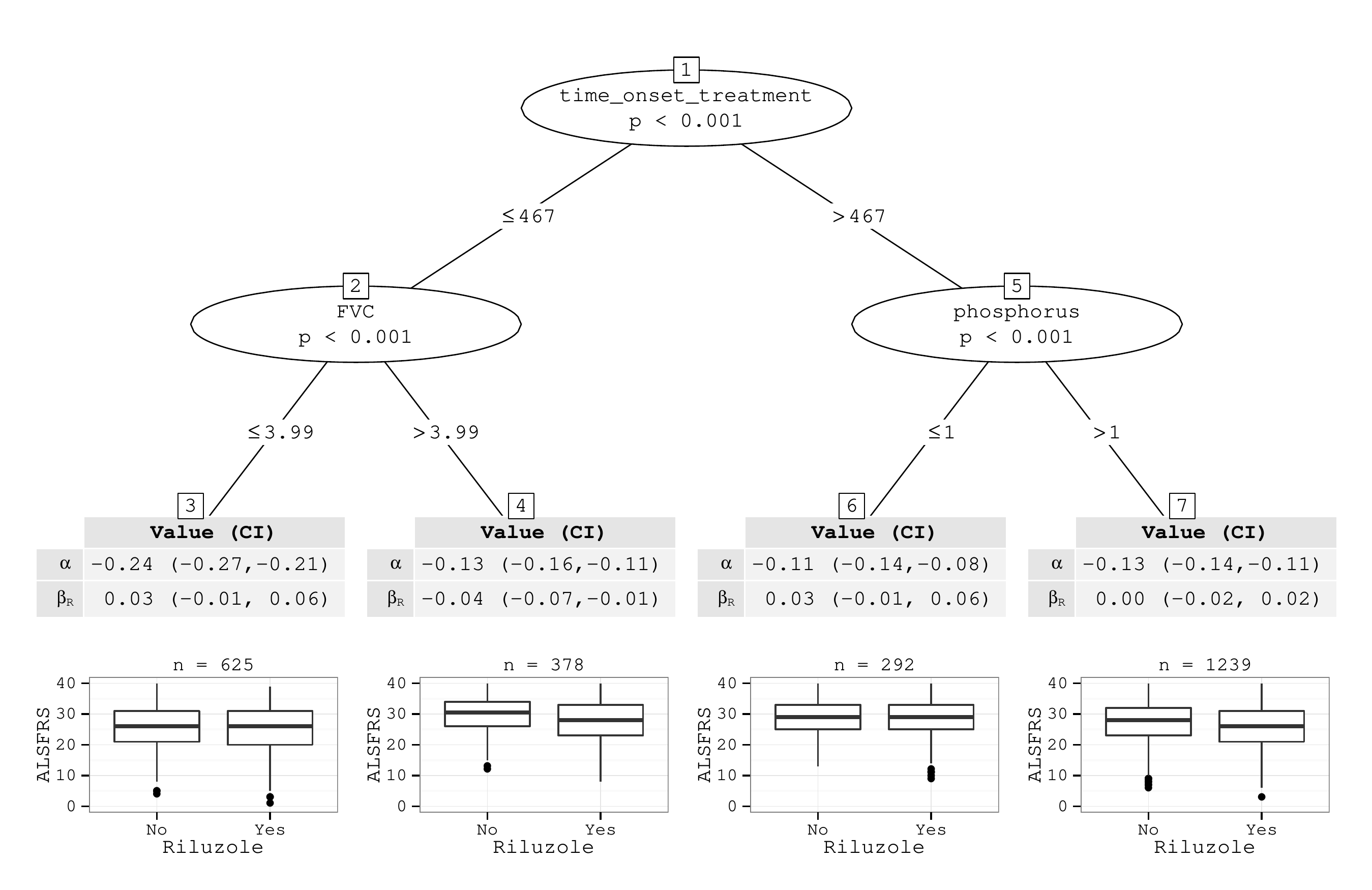}
 \caption[Result of model-based recursive partitioning with the ALSFRS as primary endpoint.]{Results of
application of model-based recursive partitioning with a Gaussian GLM with log link and offset on the data
from the PRO-ACT database with the ALSFRS score as primary endpoint variable.
Inner nodes give the split variable selected and the associated permutation test based
$p$-value for the split. 
Terminal nodes give the model coefficients including standard confidence intervals.}
 \label{ALSFRS_lmog_ctree_both1}
\end{sidewaysfigure}

The time between disease onset and start of treatment, the forced vital
capacity (FVC), and the phosphorus balance are the three partitioning
variables selected for the tree given in
Figure~\ref{ALSFRS_lmog_ctree_both1}.  The FVC value gives the volume of air
in liters that can forcibly be blown out after full inspiration to the lung. 
A normal phosphorus balance is between $1$ and $1.5$ mmol/L.  The tree
indicates a negative treatment effect of Riluzole for patients with fewer
days between disease onset and start of treatment that have a higher FVC
value (node 4).  Therefore, the FVC value is predictive in
the group of patients with less than 468 days between disease onset
and treatment start. Patients with more days between disease onset and
treatment start do not seem to have a treatment effect.
The fact that the time since onset plays an important role is not surprising
since it is a surrogate for the speed of disease progression
\citep{hothorn_randomforest4life:_2014}.  Patients with a slow progression
were seldom included early in one of the studies.  Hence a long time
between onset and start of treatment usually stands for a slow progression.

 \clearpage


\subsection{ALSFRS items}

The model for the ALSFRS sum-score assumes that the effect of Riluzole is
the same for the ten items that define the score. In a more fine-grained
analysis, we decomposed the score into its ten items (each ranging between
zero and four) and modelled each item by means of a proportional odds
model. For one of the ten items assessed at six months, \eg\ $Y_6$, the model reads
\begin{align}
  \mP(Y_6 \leq r\vert  \theox = \obsx) &= \frac{1}{1 + \exp(-\alpha_r + \beta x_R)},
\end{align}
where $r = 0,\dotsc,4$ is one of the five possible values of $Y$. The
intercept parameters are now $\bfalpha = (\alpha_0, \dots, \alpha_3)$ and the
partial score function $\psi_\alpha$ is now four dimensional.

As in the previous example, we needed to adjust for the baseline value $Y_0$,
\ie the value of the ALSFRS item read at the beginning of treatment.  This
adjustment was implemented by computing separate models; one each
for the observations with a start value $k$, which allows a baseline-specific
intercept and treatment effect :
\begin{align}
        \mP(Y_6 \leq r|Y_0 = k, X = x) &= 
	\frac{1}{1 + \exp(-\alpha_{rk} + \beta_k x_R)}
        \qquad \text{ for } k=0,\dotsc,4.
\end{align}
Therefore, we had a total of five different treatment parameters and 
$20$ different intercepts for each of the ten different items. Model-based
recursive partitioning was used to assess the parameter instability of
all $250$ parameters simultaneously. Note that some of these
parameters could not be estimated owing to too small of sample sizes; these were
simply discarded.

The implementation of the non-standard model in the theoretical and 
computational framework of model-based recursive partitioning was
straightforward. For every node, we computed the five separate models for the 
respective baseline values for each of the ten items and extracted the
partial scores. A stratified permutation test using the baseline values
as independent blocks was used to assess parameter instability. The same
procedure was applied for cut-off selection.

The resulting tree (on top of Table~\ref{items_polr_coefs}) contains splits
in time between disease onset and treatment start and in the FVC value.  The
tree is in good agreement with the tree based on the ALSFRS
(Figure~\ref{ALSFRS_lmog_ctree_both1}).  The third split variable is the
lymphocyte percentage.  Normal lymphocyte concentrations range from $16$ to
$33$ percent.  Table~\ref{items_polr_coefs} shows the coefficient values of
the models in the terminal nodes for every item and every starting value of
the given item.  Empty fields indicate that it was not possible to compute
the model.  Obviously, there were not enough observations in models with
zero as starting value for any items in any nodes.  The colours in the table
indicate whether the effect of Riluzole was positive (blue), negative (pink)
or zero (grey).  The colours were assigned on the basis of confidence
intervals of the coefficient in the given model.  Riluzole had a positive
effect on patients in the partition of terminal node $3$ who had a starting
value of $4$ in item $1$ (speech), $3$ (swallowing) or $9$ (climbing stairs)
and on patients in the partition of terminal node $7$ that had a starting
value of $3$ in item $5$ (cutting food and handling utensils).  Patients in
node 4 who had a starting value of $3$ in item $6$ (dressing and hygiene)
had a negative effect of Riluzole.  Riluzole had no effect on patients in
the partition of node $6$ which are the patients with more than 584 days
between disease onset and treatment start who have a lymphocyte
concentration under $21.5$ percent.

\begin{table}[!tbp]
{
\fontsize{7}{7.7}\selectfont 
  \begin{center}
\begin{tabular}{lccrrrr}
&& \multicolumn{5}{p{0.4\textwidth}}{\includegraphics[width=0.73\textwidth]{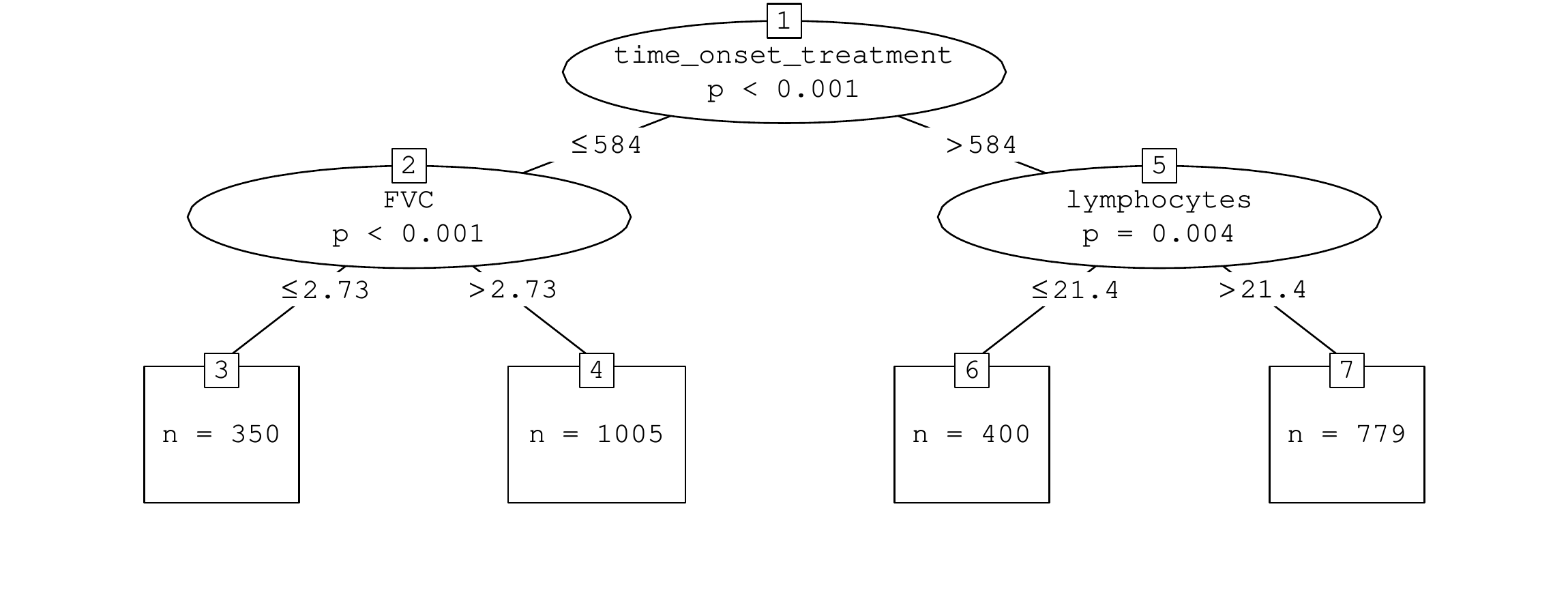}}\\[-2.5em]
			\hline \hline
\multicolumn{1}{c}{Item}&\multicolumn{1}{c}{No.}&\multicolumn{1}{c}{Start}&\multicolumn{1}{c}{Node 3}&\multicolumn{1}{c}{Node 4}&\multicolumn{1}{c}{Node 6}&\multicolumn{1}{c}{Node 7}\tabularnewline
\hline
&&&&&&\tabularnewline
\cellcolor{white}   Speech&\cellcolor{white}   1&\cellcolor{white}   $0$&\cellcolor{white}   &\cellcolor{white}   &\cellcolor{white}   &\cellcolor{white}   \tabularnewline
\cellcolor{white}   &\cellcolor{white}   &\cellcolor{white}   $1$&\cellcolor{white}   &\cellcolor{white}   &\cellcolor{white}   &\cellcolor{white}   \tabularnewline
\cellcolor{white}   &\cellcolor{white}   &\cellcolor{white}   $2$&\cellcolor{white}   &\cellcolor{gray!20}   -0.27 (-1.15,  0.60)&\cellcolor{white}   &\cellcolor{white}   \tabularnewline
\cellcolor{white}   &\cellcolor{white}   &\cellcolor{white}   $3$&\cellcolor{gray!20}    0.33 (-0.33, 1.00)&\cellcolor{gray!20}    0.04 (-0.40,  0.47)&\cellcolor{gray!20}   -0.27 (-1.12, 0.56)&\cellcolor{gray!20}   -0.06 (-0.60, 0.48)\tabularnewline
\cellcolor{white}   &\cellcolor{white}   &\cellcolor{white}   $4$&\cellcolor{blue!40}    0.84 ( 0.08, 1.59)&\cellcolor{gray!20}    0.11 (-0.28,  0.48)&\cellcolor{white}   &\cellcolor{white}   \tabularnewline
\hline
&&&&&&\tabularnewline
\cellcolor{white}   Salivation&\cellcolor{white}   2&\cellcolor{white}   $0$&\cellcolor{white}   &\cellcolor{white}   &\cellcolor{white}   &\cellcolor{white}   \tabularnewline
\cellcolor{white}   &\cellcolor{white}   &\cellcolor{white}   $1$&\cellcolor{white}   &\cellcolor{white}   &\cellcolor{white}   &\cellcolor{white}   \tabularnewline
\cellcolor{white}   &\cellcolor{white}   &\cellcolor{white}   $2$&\cellcolor{gray!20}    0.22 (-0.94, 1.40)&\cellcolor{gray!20}    0.43 (-0.48,  1.36)&\cellcolor{gray!20}    1.36 (-0.31, 3.11)&\cellcolor{gray!20}   -0.20 (-1.28, 0.88)\tabularnewline
\cellcolor{white}   &\cellcolor{white}   &\cellcolor{white}   $3$&\cellcolor{gray!20}    0.15 (-0.57, 0.87)&\cellcolor{gray!20}   -0.26 (-0.76,  0.23)&\cellcolor{gray!20}   -0.24 (-0.96, 0.47)&\cellcolor{gray!20}   -0.05 (-0.58, 0.48)\tabularnewline
\cellcolor{white}   &\cellcolor{white}   &\cellcolor{white}   $4$&\cellcolor{gray!20}    0.49 (-0.11, 1.07)&\cellcolor{gray!20}   -0.03 (-0.39,  0.32)&\cellcolor{white}   &\cellcolor{white}   \tabularnewline
\hline
&&&&&&\tabularnewline
\cellcolor{white}   Swallowing&\cellcolor{white}   3&\cellcolor{white}   $0$&\cellcolor{white}   &\cellcolor{white}   &\cellcolor{white}   &\cellcolor{white}   \tabularnewline
\cellcolor{white}   &\cellcolor{white}   &\cellcolor{white}   $1$&\cellcolor{white}   &\cellcolor{white}   &\cellcolor{white}   &\cellcolor{white}   \tabularnewline
\cellcolor{white}   &\cellcolor{white}   &\cellcolor{white}   $2$&\cellcolor{gray!20}    0.35 (-0.62, 1.32)&\cellcolor{gray!20}   -0.89 (-2.06,  0.21)&\cellcolor{gray!20}    1.51 (-0.33, 3.51)&\cellcolor{gray!20}   -0.75 (-1.96, 0.43)\tabularnewline
\cellcolor{white}   &\cellcolor{white}   &\cellcolor{white}   $3$&\cellcolor{gray!20}    0.57 (-0.06, 1.21)&\cellcolor{gray!20}   -0.36 (-0.85,  0.12)&\cellcolor{gray!20}   -0.40 (-1.17, 0.36)&\cellcolor{gray!20}    0.35 (-0.22, 0.93)\tabularnewline
\cellcolor{white}   &\cellcolor{white}   &\cellcolor{white}   $4$&\cellcolor{blue!40}    0.62 ( 0.01, 1.23)&\cellcolor{white}   &\cellcolor{gray!20}    0.15 (-0.49, 0.75)&\cellcolor{gray!20}    0.28 (-0.22, 0.75)\tabularnewline
\hline
&&&&&&\tabularnewline
\cellcolor{white}   Handwriting&\cellcolor{white}   4&\cellcolor{white}   $0$&\cellcolor{white}   &\cellcolor{white}   &\cellcolor{white}   &\cellcolor{gray!20}   -1.45 (-3.21, 0.37)\tabularnewline
\cellcolor{white}   &\cellcolor{white}   &\cellcolor{white}   $1$&\cellcolor{white}   &\cellcolor{gray!20}   -1.15 (-2.54,  0.20)&\cellcolor{gray!20}   -0.36 (-1.93, 1.25)&\cellcolor{gray!20}   -0.08 (-1.26, 1.12)\tabularnewline
\cellcolor{white}   &\cellcolor{white}   &\cellcolor{white}   $2$&\cellcolor{white}   &\cellcolor{gray!20}   -0.54 (-1.33,  0.25)&\cellcolor{white}   &\cellcolor{gray!20}    0.04 (-0.71, 0.79)\tabularnewline
\cellcolor{white}   &\cellcolor{white}   &\cellcolor{white}   $3$&\cellcolor{gray!20}   -0.13 (-0.78, 0.51)&\cellcolor{gray!20}    0.14 (-0.22,  0.49)&\cellcolor{gray!20}   -0.08 (-0.70, 0.52)&\cellcolor{gray!20}   -0.28 (-0.74, 0.17)\tabularnewline
\cellcolor{white}   &\cellcolor{white}   &\cellcolor{white}   $4$&\cellcolor{gray!20}   -0.10 (-0.71, 0.49)&\cellcolor{gray!20}    0.04 (-0.34,  0.42)&\cellcolor{white}   &\cellcolor{gray!20}   -0.14 (-0.65, 0.36)\tabularnewline
\hline
&&&&&&\tabularnewline
\cellcolor{white}   Cutting&\cellcolor{white}   5&\cellcolor{white}   $0$&\cellcolor{white}   &\cellcolor{white}   &\cellcolor{white}   &\cellcolor{white}   \tabularnewline
\cellcolor{white}   &\cellcolor{white}   &\cellcolor{white}   $1$&\cellcolor{white}   &\cellcolor{white}   &\cellcolor{gray!20}   -0.01 (-1.19, 1.15)&\cellcolor{gray!20}   -0.79 (-1.68, 0.07)\tabularnewline
\cellcolor{white}   &\cellcolor{white}   &\cellcolor{white}   $2$&\cellcolor{white}   &\cellcolor{gray!20}    0.15 (-0.54,  0.85)&\cellcolor{white}   &\cellcolor{gray!20}    0.48 (-0.14, 1.12)\tabularnewline
\cellcolor{white}   &\cellcolor{white}   &\cellcolor{white}   $3$&\cellcolor{gray!20}   -0.03 (-0.76, 0.70)&\cellcolor{gray!20}   -0.07 (-0.44,  0.30)&\cellcolor{gray!20}    0.10 (-0.60, 0.79)&\cellcolor{blue!40}    0.52 ( 0.03, 1.02)\tabularnewline
\cellcolor{white}   &\cellcolor{white}   &\cellcolor{white}   $4$&\cellcolor{gray!20}    0.13 (-0.45, 0.72)&\cellcolor{gray!20}   -0.09 (-0.49,  0.31)&\cellcolor{gray!20}   -0.14 (-0.82, 0.53)&\cellcolor{gray!20}   -0.21 (-0.74, 0.30)\tabularnewline
\hline
&&&&&&\tabularnewline
\cellcolor{white}   Hygiene&\cellcolor{white}   6&\cellcolor{white}   $0$&\cellcolor{white}   &\cellcolor{white}   &\cellcolor{white}   &\cellcolor{white}   \tabularnewline
\cellcolor{white}   &\cellcolor{white}   &\cellcolor{white}   $1$&\cellcolor{white}   &\cellcolor{white}   &\cellcolor{white}   &\cellcolor{white}   \tabularnewline
\cellcolor{white}   &\cellcolor{white}   &\cellcolor{white}   $2$&\cellcolor{white}   &\cellcolor{gray!20}   -0.11 (-0.65,  0.43)&\cellcolor{white}   &\cellcolor{gray!20}   -0.37 (-0.89, 0.15)\tabularnewline
\cellcolor{white}   &\cellcolor{white}   &\cellcolor{white}   $3$&\cellcolor{gray!20}   -0.22 (-0.88, 0.44)&\cellcolor{magenta!40}   -0.37 (-0.72,-0.03)&\cellcolor{gray!20}    0.14 (-0.50, 0.78)&\cellcolor{gray!20}    0.27 (-0.17, 0.71)\tabularnewline
\cellcolor{white}   &\cellcolor{white}   &\cellcolor{white}   $4$&\cellcolor{gray!20}    0.26 (-0.40, 0.92)&\cellcolor{gray!20}    0.01 (-0.42,  0.44)&\cellcolor{gray!20}    0.14 (-0.71, 0.98)&\cellcolor{gray!20}    0.30 (-0.31, 0.90)\tabularnewline
\hline
&&&&&&\tabularnewline
\cellcolor{white}   Bed&\cellcolor{white}   7&\cellcolor{white}   $0$&\cellcolor{white}   &\cellcolor{white}   &\cellcolor{white}   &\cellcolor{white}   \tabularnewline
\cellcolor{white}   &\cellcolor{white}   &\cellcolor{white}   $1$&\cellcolor{white}   &\cellcolor{white}   &\cellcolor{white}   &\cellcolor{white}   \tabularnewline
\cellcolor{white}   &\cellcolor{white}   &\cellcolor{white}   $2$&\cellcolor{white}   &\cellcolor{white}   &\cellcolor{gray!20}   -0.03 (-0.96, 0.89)&\cellcolor{gray!20}    0.29 (-0.47, 1.04)\tabularnewline
\cellcolor{white}   &\cellcolor{white}   &\cellcolor{white}   $3$&\cellcolor{gray!20}    0.15 (-0.57, 0.87)&\cellcolor{gray!20}   -0.32 (-0.71,  0.08)&\cellcolor{gray!20}   -0.12 (-0.74, 0.49)&\cellcolor{gray!20}   -0.05 (-0.45, 0.35)\tabularnewline
\cellcolor{white}   &\cellcolor{white}   &\cellcolor{white}   $4$&\cellcolor{gray!20}   -0.21 (-0.80, 0.36)&\cellcolor{gray!20}   -0.10 (-0.45,  0.24)&\cellcolor{gray!20}   -0.11 (-0.81, 0.57)&\cellcolor{gray!20}   -0.35 (-0.90, 0.18)\tabularnewline
\hline
&&&&&&\tabularnewline
\cellcolor{white}   Walking&\cellcolor{white}   8&\cellcolor{white}   $0$&\cellcolor{white}   &\cellcolor{white}   &\cellcolor{white}   &\cellcolor{white}   \tabularnewline
\cellcolor{white}   &\cellcolor{white}   &\cellcolor{white}   $1$&\cellcolor{white}   &\cellcolor{white}   &\cellcolor{white}   &\cellcolor{white}   \tabularnewline
\cellcolor{white}   &\cellcolor{white}   &\cellcolor{white}   $2$&\cellcolor{white}   &\cellcolor{white}   &\cellcolor{gray!20}    0.48 (-0.22, 1.16)&\cellcolor{gray!20}   -0.04 (-0.56, 0.46)\tabularnewline
\cellcolor{white}   &\cellcolor{white}   &\cellcolor{white}   $3$&\cellcolor{white}   &\cellcolor{gray!20}    0.11 (-0.33,  0.55)&\cellcolor{white}   &\cellcolor{gray!20}    0.46 (-0.12, 1.04)\tabularnewline
\cellcolor{white}   &\cellcolor{white}   &\cellcolor{white}   $4$&\cellcolor{gray!20}    0.51 (-0.16, 1.18)&\cellcolor{gray!20}    0.13 (-0.27,  0.52)&\cellcolor{white}   &\cellcolor{white}   \tabularnewline
\hline
&&&&&&\tabularnewline
\cellcolor{white}   Stairs&\cellcolor{white}   9&\cellcolor{white}   $0$&\cellcolor{white}   &\cellcolor{white}   &\cellcolor{white}   &\cellcolor{white}   \tabularnewline
\cellcolor{white}   &\cellcolor{white}   &\cellcolor{white}   $1$&\cellcolor{white}   &\cellcolor{gray!20}   -0.02 (-0.49,  0.45)&\cellcolor{gray!20}   -0.01 (-0.72, 0.68)&\cellcolor{gray!20}   -0.39 (-0.89, 0.10)\tabularnewline
\cellcolor{white}   &\cellcolor{white}   &\cellcolor{white}   $2$&\cellcolor{white}   &\cellcolor{white}   &\cellcolor{gray!20}   -0.80 (-2.10, 0.48)&\cellcolor{gray!20}    0.07 (-0.97, 1.12)\tabularnewline
\cellcolor{white}   &\cellcolor{white}   &\cellcolor{white}   $3$&\cellcolor{white}   &\cellcolor{gray!20}    0.26 (-0.19,  0.72)&\cellcolor{gray!20}   -0.65 (-1.46, 0.15)&\cellcolor{gray!20}   -0.16 (-0.79, 0.46)\tabularnewline
\cellcolor{white}   &\cellcolor{white}   &\cellcolor{white}   $4$&\cellcolor{blue!40}    1.01 ( 0.27, 1.77)&\cellcolor{gray!20}    0.06 (-0.35,  0.48)&\cellcolor{gray!20}    0.72 (-0.11, 1.55)&\cellcolor{gray!20}    0.29 (-0.32, 0.89)\tabularnewline
\hline
&&&&&&\tabularnewline
\cellcolor{white}   Respiratory&\cellcolor{white}   10&\cellcolor{white}   $0$&\cellcolor{white}   &\cellcolor{white}   &\cellcolor{white}   &\cellcolor{white}   \tabularnewline
\cellcolor{white}   &\cellcolor{white}   &\cellcolor{white}   $1$&\cellcolor{white}   &\cellcolor{white}   &\cellcolor{white}   &\cellcolor{white}   \tabularnewline
\cellcolor{white}   &\cellcolor{white}   &\cellcolor{white}   $2$&\cellcolor{white}   &\cellcolor{white}   &\cellcolor{white}   &\cellcolor{white}   \tabularnewline
\cellcolor{white}   &\cellcolor{white}   &\cellcolor{white}   $3$&\cellcolor{white}   &\cellcolor{white}   &\cellcolor{white}   &\cellcolor{white}   \tabularnewline
\cellcolor{white}   &\cellcolor{white}   &\cellcolor{white}   $4$&\cellcolor{gray!20}    0.58 (-0.06, 1.24)&\cellcolor{gray!20}   -0.08 (-0.44,  0.28)&\cellcolor{white}   &\cellcolor{white}   \tabularnewline
\hline
\end{tabular}

\caption[Coefficient of Riluzole use in the terminal nodes for every item and
every starting value in the model-based recursive partitioning with a
proportional odds model (ALSFRS items as outcome).]{Coefficient and confidence
interval of Riluzole use in the terminal nodes for every item and every
starting value in the model-based recursive partitioning with a proportional
odds model (ALSFRS items as outcome). Blue indicates a positive effect of
Riluzole, pink a negative effect and grey no
effect.\label{items_polr_coefs}}\end{center}}
\end{table}


\subsection{Survival time}

We used both a Weibull model and a Cox model to identify subgroups
with differing effects of Riluzole on the survival endpoint. The application
of the model-based recursive partitioning framework in the Weibull model
is straightforward and was introduced by 
\cite{zeileis_model-based_2008}.
Since the Cox model is a semiparametric model, where the intercept is
a function of time, treated as a nuisance parameter omitted in the 
partial likelihood, there is no direct way of obtaining $\psi_\alpha$.
Because, conceptually, deviance residuals are always defined as the derivative
of the log-likelihood with respect to the intercept, we applied martingale
residuals as $\psi_\alpha$. Also worth noting is that both models assume
proportional hazards. For the segmented model, proportional hazards are only
assumed within each partition. This has to be kept in mind when interpreting 
the treatment effect in different nodes: Parameters with different signs 
are clearly linked to opposing treatment effects, but when the parameters 
only differ in size, it is hard to say whether it is because the groups 
differ in treatment effect or because they differ in the hazard function.

\subsubsection{Weibull model}

The Weibull model is a transformation model of the form
\begin{align}
  \mP(Y \le y | X = x) = F\left(\frac{\log(y) - \alpha_1 - \beta x_R}
                                         {\alpha_2}\right),
\end{align}  
where $F$ is the cumulative distribution function of the Gompertz distribution. 
Wei\-bull models are fitted via maximum-likelihood estimation, and therefore the 
objective function in this case is the negative log-likelihood and the score function 
has one column per parameter, i.e.\ intercept $\alpha_1$, slope parameter $\beta$
and scale parameter $\alpha_2$.
In the Weibull model, we take the usual intercept as well as the scale
parameter as ``intercept''-parameter $\bfalpha = (\alpha_1, \alpha_2)^\top$
because they define the shape of the baseline hazard and hence in some
respect take the role of an intercept.  Splitting in the intercept or scale
parameter score suggests non-proportional hazards.

\begin{sidewaysfigure}
 \centering
 \includegraphics[width=\textwidth]{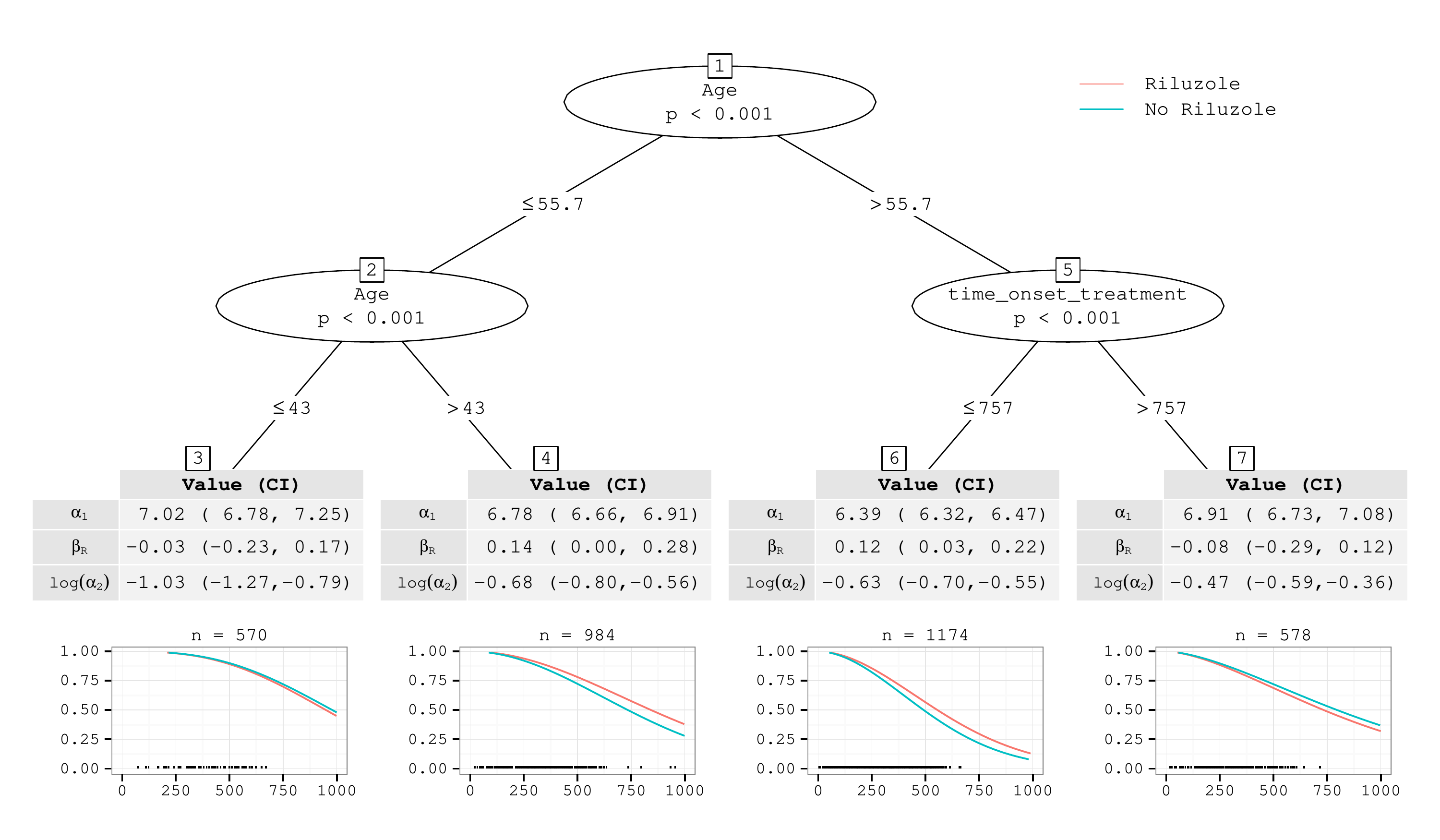}
  \caption[Result of model-based recursive partitioning using the Weibull model 
  with the survival time as
primary endpoint.]{Results of application of model-based recursive partitioning 
with a Weibull model and
data from the PRO-ACT database with survival time as primary endpoint
variable. Inner nodes give the split variable selected and the associated 
permutation test based $p$-value for the split. Terminal nodes give
the model coefficients, including standard confidence intervals and the survival curves
in the two groups of treatment. Rugs indicate event times.}
 \label{survival_wb_ctree_both}
\end{sidewaysfigure}

Figure~\ref{survival_wb_ctree_both} shows that the patient's age and again
the time between onset and treatment start play a role in the partitioning.
Older patients ($>55.7$ years) for whom the time between onset and
treatment was longer than $757$ days and very young patients did not seem to
benefit at all from the treatment. In the remaining two groups, life 
expectancy seemed to be prolonged for patients treated with Riluzole.

\subsubsection{Cox model}

The use of the Cox model in model-based recursive partitioning is a rather
special case, since the baseline hazard in the Cox model is treated as an
infinite-dimensional nuisance parameter and estimation is performed by 
minimisation of the negative partial log-likelihood.  
The Cox proportional hazards model is given by
\begin{align}
  \lambda(y|\bfx) = \lambda_0(y) \exp(\beta x_R),
\end{align}
where $\lambda$ is the hazard function and $\lambda_0$ the baseline hazard
function.  The partial score function $\psi_\alpha$ (or better,
$\psi_{\lambda_0}$) cannot be easily derived.  As surrogate score function,
we propose using the martingale residuals as a score for the baseline hazard,
which takes the role of an intercept in the Cox model, and the score
residuals for the treatment parameters $\beta$.  The score residuals are an
intuitive choice because they are the first derivative of the partial
log-likelihood with respect to the parameters.
We used martingale residuals to check whether there is a
general difference in the endpoint for different patients, which in
parametric models is usually shown by the score with respect to the
intercept.  Instability in the martingale residuals 
indicates a violation of the proportional hazards assumption.  Since
the martingale residuals are not normally distributed, the application of
permutation tests is more appropriate than the use of M-fluctuation tests.

Age and the time between disease onset and start of Riluzole treatment form
the segments in this example.  The tree in this example has almost the same
splits as the tree in the previous example.  Also estimates support the
results of the Weibull example.  Again, we did not see much difference
between treated and untreated very young patients.  For all other groups,
Riluzole treatment led to a slight tendency for a lower risk of death.

\begin{sidewaysfigure}
 \centering
 \includegraphics[width=\textwidth]{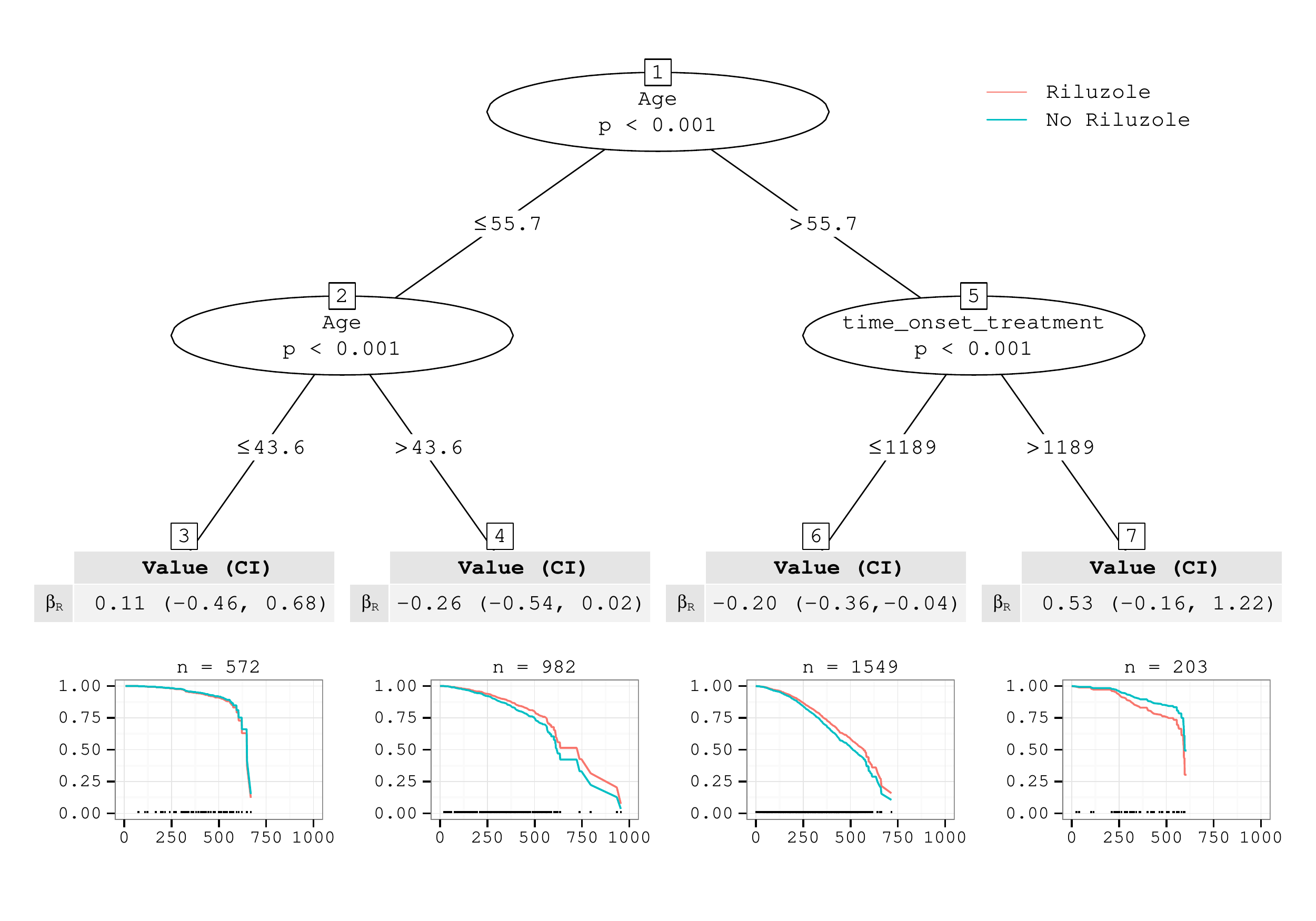}
 \caption[Result of model-based recursive partitioning using the Cox model 
 with the survival time as primary endpoint.]{Results of application of 
 model-based recursive partitioning with a Cox model and data from the
PRO-ACT database with the survival time as primary endpoint variable. Inner
nodes give the split variable selected and the associated permutation test 
based $p$-value for the split. Terminal nodes give the
model coefficients including confidence intervals and the survival curves in
the two groups of treatment. Rugs indicate event times.}
 \label{survival_cox_ctree_both1}
\end{sidewaysfigure}

\clearpage

\section{Discussion}\label{dis}
 
Model-based recursive partitioning allows the direct segmentation of the model 
describing the overall treatment effect as specified in the study protocol.
This is the most important benefit of embedding subgroup analysis into this
framework because it would be hard to explain why the overall treatment
effect and the partitioned treatment effect have to be estimated by two
different procedures. This renders the application of suboptimal models unnecessary, 
such as when a change score is analysed using linear models 
\citep{dusseldorp_qualitative_2013}.

Although we are conceptually only interested in finding predictive factors,
we think it is necessary to allow splits in the partial scores with respect
to both intercept and treatment parameter. This procedure will also detect
prognostic factors, but there is a higher chance of including all relevant
predictive factors since one might miss prognostic factors when only 
the treatment scores are split. In our analysis, we decided on the nature of the
partitioning variables (prognostic or predictive) only when we interpreted the
results of the analysis.

In a model with more covariates than the treatment (e.g. strata), we 
would still split the partial scores with respect to intercept and treatment
parameter for subgroup
analyses. A theoretical assumption is then that the parameters that are
not split stay constant. In practice, this assumption usually does not 
hold. It is generally also possible to split more than just the scores
with respect to intercept and treatment parameter. Then the split variables
are not restricted to being predictive or prognostic but may have an
association with the effect of the other covariates.

In model-based recursive partitioning, the variable selection in each node is
error controlled, \ie the probability of selecting a partitioning variable
for splitting, when actually all partitioning variables are independent of
the scores, is at most as large as the nominal level. 
The only drawback of using multiple testing procedures is in cases where 
there are many possible partitioning variables that do not contain information,
because with increasing number of noise variables the chance of detecting an 
actually existing subgroup goes down.
The application of
permutation tests has the advantage of taking the correlation structure
among the partitioning variables into account. Furthermore, for small
studies or small subgroups, the exact conditional $p$-value can be easily
approximated up to any desired accuracy; therefore, the method does not rely on
asymptotic arguments. The trees obtained by model-based recursive
partitioning allow straightforward visualisation, potentially enriched with
plots illustrating the distribution of the endpoints for the different
treatment groups in each subgroup. Therefore, the results of such a subgroup
analysis are easily communicated to physicians. Looking at a tree is much
easier than trying to understand the meaning of higher-order interactions in
a linear predictor.
A general drawback of tree methods is the instability of the tree structure
with respect to small perturbations in the data, whereas the resulting
partitions we are primarily interested in are often relatively stable
\citep{hothorn_unbiased_2006}.  Instability in the tree structure can be
assessed by means of the variable selection and split statistics, where it is
easy to identify all equally likely splits. Bootstrap aggregation and forest
procedures are well-known for their ability to stabilise single trees
\citep{strobl_introduction_2009} at the cost of interpretability  and point
into a promising future research direction also for model-based recursive
partitioning.

The statistical properties of the confidence intervals derived from the
segmented model await further attention. 
\cite{leeb_model_2005} discuss the validity of inference after variable
selection and claim that it is difficult if at all possible.
\cite{bai_computation_2003}, who discuss the construction of confidence
intervals after splitting up the data based on a break point in a single
partitioning variable, argue that it is possible.  In our approach we first
search for the most appropriate partitioning variable (variable selection) and
then search for the optimal split point (break point selection). To our
knowledge there is no literature on inference after variable and break point
selection and thus it is unclear if or how valid confidence intervals can be
computed.
In any case  the results of such a subgroup analysis have to be confirmed in
follow-up trials, which lowers the necessity of confidence intervals. To be
conservative one can see the confidence intervals for parameters in the
subgroup-specific models as shown in our examples as a range of possible values
and hence as a measure of variability rather than significance
\citep{lagakos_challenge_2006}.


It would be interesting to extend the framework of the PRO-ACT database of ALS
studies to models for non-independent data, such as mixed models for
longitudinal observations. This would allow ALS disease progression to be
modelled over time, and also a potentially time-varying treatment effect to be
assessed.  In our way of modelling the functional endpoint, we include no
information about patients that died within the first six months after
treatment start.  Joint modelling of the longitudinal functional endpoint and
the survival endpoint is a means of combining all possible information
\citep{henderson_joint_2000}.

Despite the deficits of model-based recursive partitioning for subgroup
analysis discussed in this section, we think that the procedure as
introduced and illustrated in this paper rather closely resembles the
requirements for statistical procedures in this field as outlined in the EMA
guideline \citep{ema_ema_2014}.  In particular, it is the most generally
applicable procedure with statistical error control and unbiased variable
selection \citep{hothorn_unbiased_2006, zeileis_model-based_2008}.  With
the available open-source implementation (see following section for
details), the method can be applied straightforwardly elsewhere.

\section*{Computational details}

An open-source implementation of all methods discussed in this paper and beyond
is available in the \textbf{partykit} package \citep{hothorn_partykit:_2014}.
PRO-ACT data are available at \url{https://nctu.partners.org/ProACT/}
\citep{massachusetts_general_hospital_pooled_2013}.  The source code for
reading and cleaning the database is provided in the \textbf{TH.data} package
\citep{hothorn_data_2014}. The source code for the analyses is provided in the
supplementary material. All computations were conducted using \textbf{partykit}
(version 0.8-2) in the \textsf{R} system for statistical computing
\citep[][version 3.1.2]{R}.

{\footnotesize
\begin{lstlisting}[language=R, caption=Code snippet for Weibull model in model-based
recursive partitioning using the function \texttt{ctree()} from the \textbf{partykit} package.]
## Function to compute Weibull model and return score matrix
mywb <- function(data, weights, parm) {  
  mod <- survreg(Surv(survival.time, cens) ~ Riluzole, 
                 data = data, subset = weights > 0, 
                 dist = "weibull")
  ef <- as.matrix(estfun(mod)[,parm])
  ret <- matrix(0, nrow = nrow(data), ncol = ncol(ef))
  ret[weights > 0,] <- ef
  ret
}

## Compute tree
tree <-  ctree(fm, data = data, ytrafo = my.wb, 
               control = ctree_control(maxdepth = 2, 
  		               testtype = "Bonferroni"))
\end{lstlisting}
}

\section*{Acknowledgements}

We are thankful to the organisers and participants of the
``Workshop on Classification and Regression Trees'' (March 2014),
sponsored by the Institute for Mathematical Sciences of the
National University of Singapore, for helpful feedback and
stimulating discussions and to Karen A.~Brune for improving the language.

\bibliography{MOB_subgroup_TR}

\end{document}